\documentclass[aps,prl,twocolumn,showpacs,preprintnumbers,amsmath,amssymb,superscriptaddress]{revtex4}%
\usepackage{graphicx}
\usepackage{dcolumn}
\usepackage{bm}

\newcommand{\YPr}{Y$_{1-x}$Pr$_{x}$Ba$_2$Cu$_3$O$_{7-\delta}$~}
\newcommand{\Y}{YBa$_2$Cu$_3$O$_{7-\delta}$ }
\newcommand{\sixO}{$^{16}$O}
\newcommand{\eightO}{$^{18}$O}

\newcommand{\Cel}{$^{o}$C }

\begin{document}
\preprint{PREPRINT (\today)}
\title{Direct observation of the oxygen isotope
effect on the in-plane magnetic field penetration depth in
optimally doped YBa$_2$Cu$_3$O$_{7-\delta}$}
\author{R.~Khasanov}
\email{rustem.khasanov@psi.ch} \affiliation{Physik-Institut der
Universit\"{a}t Z\"{u}rich, Winterthurerstrasse 190, CH-8057,
Switzerland} \affiliation{Laboratory for Muon Spin Spectroscopy,
PSI, CH-5232 Villigen PSI, Switzerland}
\author{D.G.~Eshchenko}
\affiliation{Physik-Institut der Universit\"{a}t Z\"{u}rich,
Winterthurerstrasse 190, CH-8057, Switzerland}
\affiliation{Laboratory for Muon Spin Spectroscopy, PSI, CH-5232
Villigen PSI, Switzerland}
\author{H.~Luetkens}
\affiliation{Laboratory for Muon Spin Spectroscopy, PSI,
CH-5232 Villigen PSI, Switzerland}
\affiliation{Institut f\"{u}r Metallphysik und Nukleare
Festk\"{o}rperphysik, TU Braunschweig, 38106 Braunschweig, Germany}
\author{E.~Morenzoni}
\affiliation{Laboratory for Muon Spin Spectroscopy, PSI,
CH-5232 Villigen PSI, Switzerland}
\author{T.~Prokscha}
\affiliation{Laboratory for Muon Spin Spectroscopy, PSI,
CH-5232 Villigen PSI, Switzerland}
\author{A.~Suter}
\affiliation{Laboratory for Muon Spin Spectroscopy, PSI,
CH-5232 Villigen PSI, Switzerland}
\author{N.~Garifianov}
\affiliation{Laboratory for Muon Spin Spectroscopy, PSI, CH-5232
Villigen PSI, Switzerland}

\author{M.~Mali}
\affiliation{Physik-Institut der Universit\"{a}t Z\"{u}rich,
Winterthurerstrasse 190, CH-8057, Switzerland}
\author{J.~Roos}
\affiliation{Physik-Institut der Universit\"{a}t Z\"{u}rich,
Winterthurerstrasse 190, CH-8057, Switzerland}
\author{K.~Conder}
\affiliation{Laboratory for Neutron Scattering, ETH Z\"urich and
PSI, CH-5232 Villigen PSI, Switzerland}
\author{H.~Keller}
\affiliation{Physik-Institut der Universit\"{a}t Z\"{u}rich,
Winterthurerstrasse 190, CH-8057, Switzerland}

\begin{abstract}
We report the first {\it direct} observation of the oxygen-isotope
($^{16}$O/$^{18}$O) effect on the in-plane penetration depth
$\lambda_{ab} $ in a nearly optimally doped
YBa$_2$Cu$_3$O$_{7-\delta}$ film using the novel low-energy
muon-spin rotation technique.  Spin polarized low energy muons are
implanted in the film at a known depth $z$ beneath the surface and
precess in the local magnetic field $B(z)$.  This feature allows
us to measure directly the profile $B(z)$ of the magnetic field
inside the superconducting film in the Meissner state and to make
a model independent determination of $\lambda_{ab}$. A substantial
isotope shift $\Delta\lambda_{ab}/\lambda_{ab}=2.8(7)$\% at 4~K is
observed, implying that the in-plane effective supercarrier mass
$m_{ab}^\ast$ is oxygen-isotope dependent with $\Delta
m_{ab}^\ast/m_{ab}^\ast = 5.5(1.4)\%$.
\end{abstract}
\maketitle

One of the fundamental questions concerning the physics of cuprate
high-temperature superconductors (HTSC) is whether the
electron-phonon interaction plays an essential role in these
systems or not. The high values of the superconducting transition
temperature $T_{c}$ and the observation of only a tiny
oxygen-isotope shift of $T_{c}$ in optimally doped HTSC (see, {\it
e.g.} \cite{Franck94}) were taken as important arguments to
propose alternative pairing mechanisms of purely electronic
origin.  The conventional phonon-mediated theory of
superconductivity is based on the Migdal adiabatic approximation
in which the effective supercarrier mass $m^{\ast}$ is independent
of the mass $M$ of the lattice atoms.  However, if the interaction
between the carriers and the lattice is strong enough, the Migdal
adiabatic approximation breaks down and $m^{\ast}$ depends on $M$
(see, {\it e.g.} \cite{Alexandrov94}).  The ideal experiment to
explore a possible coupling of the supercarriers to the lattice is
an isotope effect study of the magnetic field penetration depth.
For HTSC, which are superconductors in the clean limit, the
in-plane penetration depth $\lambda_{ab}$ obeys the relation:
\begin{equation}
\label{Lambda}
1/\lambda_{ab}^{2} \propto n_{s}/m^{\ast}_{ab} \; ,
\end{equation}
where $n_{s}$ is the superconducting charge carrier density and
$m^{\ast}_{ab}$ is the in-plane effective mass of the charge
carriers. Note that there is no such simple relation for $T_{c}$.
Since $n_s$ was found to be predominantly isotope independent
\cite{Zhao95,Zhao97,Zhao98,Hofer00,Zhao01,Zhao01a,Khasanov03},
isotope effect experiments on the penetration depth turned out to
be a unique tool to investigate unconventional lattice effects in
HTSC. Previous oxygen-isotope effect studies of the penetration
depth in \Y \cite{Zhao95}, La$_{2-x}$Sr$_{x}$CuO$_{4}$
\cite{Zhao97,Zhao98,Hofer00},
Bi$_{1.6}$Pb$_{0.4}$Sr$_{2}$Ca$_{2}$Cu$_{3}$O$_{10+\delta}$~
\cite{Zhao01} and \YPr \cite{Khasanov03} showed indeed a
pronounced oxygen isotope dependence of the supercarrier mass.  In
all these experiments, however, the penetration depth was
determined {\it indirectly} from magnetization measurements
\cite{Zhao95}, the Meissner fraction \cite{Zhao97,Zhao98},
magnetic torque \cite{Hofer00}, and muon-spin rotation ($\mu$SR)
experiments \cite{Khasanov03}.  The expression "indirectly" means
that all these techniques require a model to extract the absolute
value of the penetration depth $\lambda$.  In contrast,
measurements of the exponential field decay ($B(z)=
B_0\exp(-z/\lambda)$) in the Meissner state allow to determine
$\lambda$ in a {\it direct way}. Recently, such a direct
measurement of the in-plane magnetic penetration depth
$\lambda_{ab}$ in a thin \Y film was performed \cite{Jackson00} at
the Paul Scherrer Institute (PSI, Switzerland) by using the novel
low-energy $\mu$SR (LE$\mu$SR) technique \cite{Morenzoni94}. In
this experiment  the muon implantation depth was controlled by the
variation of the incoming muon energy and the magnitude of the
field was monitored by the muon spin precession frequency, alike
in standard $\mu$SR.

In this Letter, we report on the oxygen-isotope (\sixO/\eightO)
effect (OIE) on the in-plane magnetic field penetration depth in a
high-quality \Y film near optimal doping measured directly by
LE$\mu$SR.  The experiments revealed that in the Meissner state
the magnetic field $B(z)$ in the $^{16}$O substituted sample
decreases stronger with distance $z$ than in the $^{18}$O
substituted sample, clearly demonstrating that $^{16}\lambda_{ab}
< ^{18}\lambda_{ab}$.

The samples used for the LE$\mu$SR experiments were high quality
\Y films with an area of 2x3 cm$^2$ and a thickness of 600~nm
supplied by {\it Theva} \cite{Theva}.  The films were grown by
thermal coevaporation of the constituents onto a single-crystal
barium titanate substrate.  The oxygen-isotope exchange was
performed by annealing the samples in $^{18}$O$_2$ gas.  In order
to ensure that the $^{16}$O and $^{18}$O substituted samples are
subject of the same thermal history, the annealing (in
$^{16}$O$_2$ and $^{18}$O$_2$) was performed simultaneously. The
exchange process was carried out at 600\Cel during 25~h, followed
by slow cooling (20$^o$C/h) in order to oxidize them completely.
The fraction of the $^{18}$O content in the film was estimated to
be 95\%, corresponding to the $^{18}$O content in the initial gas
used for the annealing process. The oxygen content (7-$\delta$)
and the quality of the films were estimated from 1mT field-cooled
(FC) SQUID magnetization measurements. The observed $T_c$ onsets
are $^{16}T_c= 89.3(1)$~K and $^{18}T_c=89.1(1)$~K. The transition
widths are $\simeq$0.9~K for both samples.  A comparison of the
$T_c$'s with the known experimental $T_c$ vs $\delta$ curves for
$^{16}$O and $^{18}$O substituted \Y \cite{Zech95} yields
$\delta=0.150(5)$ for both films.  While $\delta$ indicates that
the samples are slightly underdoped the narrow transition shows
that the oxygen distribution over the films is quite homogeneous.
The results of the OIE on $T_{c}$ are summarized in
Table~\ref{OIEresults}.

\begin{widetext}

\begin{table}[htb]
\caption[~]{\label{OIEresults}
Summary of the OIE results for \Y extracted from the experimental
data (see text for an explanation).
} %
\begin{center}
\begin{tabular}{llllllllll}\\ \hline
\hline Method &Sample &$^{16}T_c$&$^{18}T_c$&$\Delta T_c/
T_c$~~&$\Delta \lambda_{ab}/\lambda_{ab}$~~&$\Delta
m_{ab}^\ast/m_{ab}^\ast$ \\
&&(K)&(K)&(\%)&(\%)&(\%)\\
\hline
LE$\mu$SR&Thin film&89.3(1)&89.1(1)&-0.22(16)&2.8(7)&5.5(1.4) \\
\hline

Magnetization~~&Fine powder~~&91.66(3)&91.42(3)&-0.26(5)&3.0(1.1)&6.0(2.2) \\

&&91.71(3)\footnotemark[1]&91.45(1)\footnotemark[2]&-0.28(5)&2.4(1.0)&4.8(2.0)\\

 \hline \hline \\
\end{tabular}
\footnotetext[1]{ results for the back-exchange (\eightO $\to$
\sixO) sample }
\footnotetext[2]{ results for the back-exchange (\sixO $\to$
\eightO) sample }
\end{center}
\end{table}
\end{widetext}

The transverse-field LE$\mu$SR experiments were performed on the
$\pi$E3 muon beam line at PSI. A weak external magnetic field
$B_{0}=9.2$~mT was applied parallel to the sample surface after
the sample was cooled in zero magnetic field from a temperature
above $T_c$ to 4~K.  Spin-polarized muons were implanted at a
depth ranging from 20-150~nm beneath the surface of the film by
varying the energy of the incident muons from 3 to 30~keV. For
each implantation energy a time-differential $\mu$SR spectrum was
measured. The muon implantation depth profile $n(z)$ for the given
implantation energy was calculated using a Monte-Carlo code
TRIM.SP \cite{Eckstein92}. The reliability of the calculated
$n(z)$ was crosschecked by previous LE$\mu$SR experiments on thin
metal layers \cite{Morenzoni02}.

\begin{figure}[htb]
\centering
\includegraphics[width=0.9\linewidth,angle=-90]{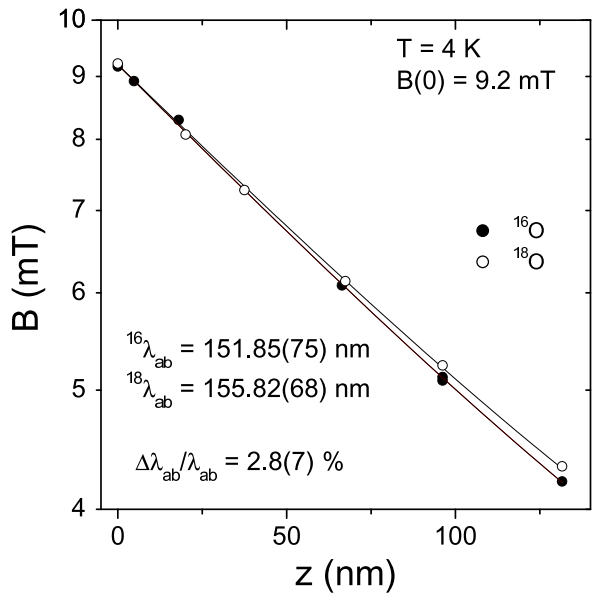}
\caption{Magnetic field penetration profiles $B(z)$ for a $^{16}$O
substituted (closed symbols) and a $^{18}$O substituted (open
symbols) \Y film measured in the Meissner state at 4~K and an
external field of 9.2~mT, applied parallel to the surface of the
film.  The data are shown for implantation energies 3, 6, 10, 16,
22, and 29~keV starting from the surface of the sample.  Solid
curves are best fits by Eq.(\ref{exp_decay_film}).
 }
\label{LEMU1}
\end{figure}

The experimental data were analyzed in the following way. For each
implantation energy the average value of the magnetic field $\bar
B$  and correspondent average value of the stopping distance $\bar
z$ were extracted.  The value of $\bar B$ was taken from the fit
of the time evolution of the muon-spin polarization spectrum by
using the Gaussian relaxation function:
\begin{equation}
\label{gauss_fit}
a(t)=a_0\exp(-\sigma^2t^2/2)\cos(\gamma \bar B t + \phi) \; ,
\end{equation}
where $a_0$ is the initial asymmetry, $\sigma$ is the Gaussian
relaxation rate, and $\gamma=2\pi\cdot 135$MHz/T is the
gyromagnetic ratio of the muon.  Note that at our level of
statistics ($\sim 5\times10^5$ events per spectrum) the fit of
experimental data with Eq.~(\ref{gauss_fit}) satisfies the
$\chi^2$ criterium.  The value of $\bar z$ was taken as the first
moment of the emulated $n(z)$ distribution.  Results of this
analysis for the $^{16}$O and $^{18}$O substituted \Y films are
shown in Fig.~\ref{LEMU1}.  The data points for the $^{18}$O film
are systematically higher than those for the $^{16}$O film,
showing that $^{16}\lambda_{ab} < ^{18}\lambda_{ab}$.  The solid
lines represent a fit to the $\bar B$ data by the function:
\begin{equation}
\label{exp_decay_film}
B(z)=B_0\frac{\cosh[(t-z)/\lambda_{ab}]}{\cosh(t/\lambda_{ab})} \; .
\end{equation}
This is the form of the classical exponential field decay in the
Meissner state $B(z)=B_0 \exp (-z/\lambda_{ab})$, modified for a
film with thickness $2t$ with flux penetrating from both sides.
The value of $z$ was corrected by $z_0=8$~nm, corresponding to a
"dead layer", which mainly arises from the surface roughness of
the film \cite{Jackson00}. Fits with Eq.~(\ref{exp_decay_film}) to
the extracted $^{16} \bar B(\bar z)$ and $^{18} \bar B(\bar z)$
yield $^{16}\lambda_{ab}({\rm 4K})=151.85(75)$~nm and
$^{18}\lambda_{ab}({\rm 4K})=155.82(68)$~nm. Taking into account a
$^{18}$O content of 95\%, the relative shift was found to be
$\Delta\lambda_{ab}/\lambda_{ab}=(^{18}\lambda_{ab}-^{16}\lambda_{ab})/^{16}\lambda_{ab}
= 2.8(7)$\% at 4~K. This value is consistent with previous results
for optimally doped \Y \cite{Zhao95} and
Bi$_{1.6}$Pb$_{0.4}$Sr$_{2}$Ca$_{2}$Cu$_{3}$O$_{10+\delta}$\cite{Zhao01}
obtained from {\it indirect} magnetization measurements.  Here we
would like to mention two important points: (i) The good agreement
between the experimental points and theoretical curves
(Eq.~(\ref{exp_decay_film})) indicates that the films are
homogenous over the whole thickness.  (ii) $\lambda_{ab}$ obtained
from LE$\mu$SR is model independent.  It is extracted from the
{\it measured} field profile in the Meissner state using only one
fit parameter $\lambda_{ab}$.

\begin{figure}[htb]
\centering
\includegraphics[width=0.9\linewidth,angle=-90]{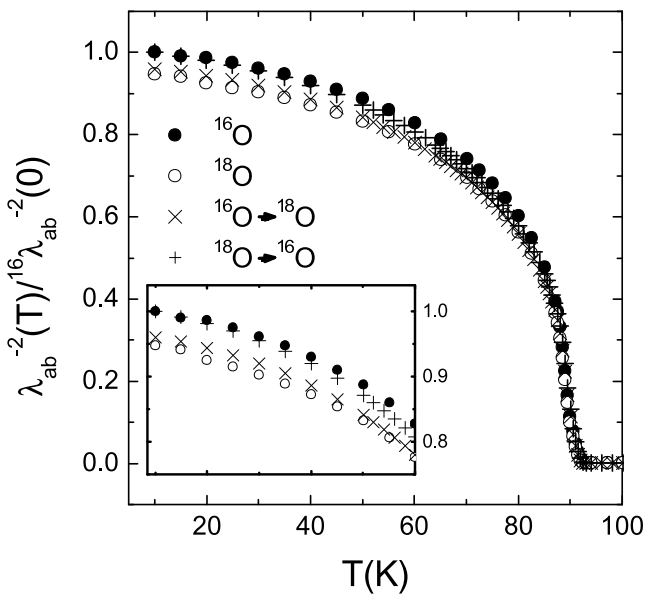}
\caption{\label{Shoenberg_data} Temperature dependence of
$\lambda_{ab}^{-2}$ normalized by $^{16}\lambda_{ab}^{-2}(0)$ for
$^{16}$O and $^{18}$O substituted \Y fine powder samples as
obtained from low-field SQUID magnetization measurements. The
inset shows the low-temperature region between 10~K and 60~K. The
reproducibility of the oxygen exchange procedure was checked by
the backexchange (crosses).} \label{magnetization}
\end{figure}

We also performed additional OIE experiments on $\lambda_{ab}$
based on measurements of the Meissner fraction in fine \Y powder.
The powder samples were ground for about 60~min and then passed
through a 10~$\mu$m sieve. The oxygen exchange procedure was
analogous to the one described above for the thin film samples.
The \eightO~ content in the sample, as determined from the change
of the sample weight after the isotope exchange, was found to be
89(2)\%.  The value of the Meissner fraction $f$ was calculated
from 1~mT FC SQUID magnetization measurements.  The absence of
weak links between grains was confirmed by the linear magnetic
field dependence of the FC magnetization measured at 5~K in
0.5~mT, 1~mT, and 1.5~mT. The temperature dependence of the
magnetic field penetration depth $\lambda_{\rm eff}$ (powder
average) was calculated from $f$ under the assumption that the
sample grains are spherical \cite{Shoenberg40}. The in-plane
penetration depth $\lambda_{ab}(T)$ (Fig.~\ref{magnetization}) was
determined from the measured $\lambda_{\rm eff}(T)$ using the
relation $\lambda_{\rm eff}=1.31\lambda_{ab}$ which holds for
highly anisotropic superconductors ($\lambda_c/\lambda_{ab}>5$)
\cite{Fesenko91}. Taking into account a $^{18}$O content of 89\%,
the relative shift at 4~K was found to be
$\Delta\lambda_{ab}/\lambda_{ab}=3.0(1.1)$\% in a good agreement
with the LE$\mu$SR data (see Table~\ref{OIEresults}).  In order to
substantiate the intrinsic character of the observed OIE on
$\lambda_{ab}$, we performed a back exchange experiment.  For this
purpose the $^{16}$O sample was annealed in $^{18}$O$_2$ gas
($^{16}$O$\to ^{18}$O) and the $^{18}$O sample in $^{16}$O$_2$ gas
($^{18}$O$\to ^{16}$O).  The annealing procedure was analogous to
the one described above for the thin film samples.  The results of
the back exchange experiments are also shown in
Fig.~\ref{Shoenberg_data} (cross symbols). All the results of the
OIE on $T_c$ and $\lambda_{ab}$ are summarized in
Table~\ref{OIEresults}.

According to Eq.~(\ref{Lambda}) the
OIE on $\lambda_{ab}$
is due to a shift in $n_{s}$ and/or $m_{ab}^{\ast}$: %
\begin{equation}
\label{DeltaLambda}
\frac{\Delta \lambda_{ab}}{\lambda_{ab}}= \frac{1}{2} \frac{\Delta
m^\ast_{ab}}{m^\ast_{ab}} - \frac{1}{2} \frac{\Delta n_s}{n_s} \; .
\end{equation}
Previous OIE studies
in HTSC's \cite{Zhao95,Zhao97,Zhao98,Hofer00,Zhao01,Zhao01a,Khasanov03}
showed a
pronounced OIE on $m^\ast_{ab}$ with negligible OIE on $n_s$.
In the present work we provide further evidence of this scenario
from measurements of the nuclear quadrupole resonance (NQR) frequency
($\nu_Q$) of the plane ($\nu_Q^{pl}$) and the chain ($\nu_Q^{ch}$)
$^{63}$Cu in $^{16}$O and $^{18}$O substituted \Y powder samples.  It
is known \cite{Imai93,Karpinski01} that in HTSC $\nu_Q^{pl}$ and
$\nu_Q^{ch}$ are very sensitive to changes in the density of mobile
carriers $n$ in the superconducting CuO$_2$ planes ($n^{pl}$) and
CuO chains ($n^{ch}$).  Both $\nu_Q^{pl}$ and $\nu_Q^{ch}$ increase
with doped hole concentration roughly 20~MHz per one doped hole per Cu
atom \cite{Stern94}.
The NQR measurements of $^{63}$Cu at 94~K show that within 5~kHz
error bar limits the plane copper and the chain copper NQR
frequencies are the same in $^{16}$O and $^{18}$O substituted \Y
powder samples (see Table~\ref{NQRresults}).  This implies that by
the oxygen substitution $n^{pl}$ and $n^{ch}$ change for less than
$3 \times 10^{-4}$ hole per Cu atom.  In optimally doped \Y one
has two plane and one chain Cu atom in the unit cell.  From the
NQR measurements we can thus conclude that the change of the hole
number per unit cell is less than $3\cdot 3 \times 10^{-4} \simeq
10^{-3}$ hole.  Since in optimally doped \Y there is approximately
one doped hole per unit cell, the relative change of hole density
$\Delta n/n$ at the oxygen exchange must be less than $10^{-3}$.
The NQR results are summarized in Table~\ref{NQRresults}.
For \Y
it was shown \cite{Fiory90} that $n_s \approx n$
at $T < 70$~K.
Based on this observation we thus can conclude that $\vert
1/2\cdot\Delta n_s/n_s\vert < 0.05\%$ in Eq.(\ref{DeltaLambda}).
Consequently, the main contribution to the OIE on $\lambda_{ab}$
has to come from the isotope dependence of the in-plane charge
carrier mass $m^\ast_{ab}$, so that
$\Delta\lambda_{ab}/\lambda_{ab} \simeq (\Delta
m^{\ast}_{ab}/m^{\ast}_{ab}$)/2.  With
$\Delta\lambda_{ab}/\lambda_{ab}=2.8(7)\%$ we obtain $\Delta
m^{\ast}_{ab}/m^{\ast}_{ab} = 5.5(1.4)\%$ at 4~K (see
Table~\ref{OIEresults}).  This result is remarkable in spite of
the fact that the observed OIE on $T_{c}$ in optimally doped \Y is
rather small ($\Delta T_{c}/T_{c} = - 0.26(5)\%$, see
Table~\ref{OIEresults}).

\begin{table}[htb]
\caption[~]{\label{NQRresults}
Results of plane and chain $^{63}$Cu NQR in $^{16}$O and $^{18}$O
substituted \Y powder samples at 94~K (see text for an explanation). } %
\begin{center}
\begin{tabular}{lllccccc} \hline\hline
&\multicolumn{2}{c}{$^{16}$O}&\  &\multicolumn{2}{c}{$^{18}$O}&&\\
\cline{2-3}
\cline{5-6}
&$\nu_Q$&Linewidth&&$\nu_Q$&Linewidth~~&$\Delta\nu_Q/\nu_Q$ \\
&(MHz)&(KHz)&&(MHz)&(KHz)&(\%)\\
\hline
Planes&31.580(5)&336(10)&&31.580(5)&340(10)&0.00(2)\\

Chains&22.050(5)&161(5)&&22.050(5)&161(5)&0.00(3)\\
\hline \hline
\end{tabular}
\end{center}
\end{table}

Note that such an isotope effect on $m^*_{ab}$ is {\em not
expected} for a conventional weak-coupling phonon-mediated BCS
superconductor.  In fact in HTSC the charge carriers are strongly
coupled to optical phonons, as demonstrated by measurements of the
static and high-frequency dielectric constants
\cite{Alexandrov00}, photoemission \cite{Shen01} and tunnelling
\cite{Ponomarev99} experiments.  Strong interaction between the
charge carriers and the lattice ions leads to a break down of the
adiabatic Migdal approximation \cite{Alexandrov94}, and
consequently the effective supercarrier mass $m^\ast$ depends on
the mass of the lattice atoms.  To our knowledge there are just a
few theoretical models which predict an OIE on the effective
carrier mass $m^*$ (see {\it e.g.}
Refs.~\onlinecite{Scalapino87,Alexandrov94,Grimaldi98,Bussmann03}).

In conclusion, we used LE$\mu$SR to measure {\it directly} the
oxygen isotope ($^{16}$O/$^{18}$O) effect on the in-plane magnetic
field penetration depth $\lambda_{ab}$ in optimally doped \Y
films. The OIE on $\lambda_{ab}$ at 4~K was found to be
$\Delta\lambda_{ab}/\lambda_{ab}=2.8(7)$\%. The intrinsic
character of the OIE on $\lambda_{ab}$ was confirmed by back
exchange experiments on fine powders with low-field SQUID
magnetization measurements.  It is concluded that the OIE arises
mainly from the isotope dependence of the in-plane charge carrier
mass $m^*_{ab}$ with $\Delta m_{ab}^\ast/m_{ab}^\ast = 5.5(1.4)\%$
at 4~K. This finding implies that even in optimally doped cuprate
superconductors for which only a small isotope effect on $T_{c}$
is observed, the supercarriers are strongly coupled to the
lattice.

This work was partly performed at the Swiss Muon Source (S$\mu$S)
at the Paul Scherrer Institute (Villigen, Switzerland). The
authors are grateful to A.~Bussmann-Holder, K.A.~M\"uller,
T.~Schneider, and Z.X.~Shen for stimulating discussions.  This
work was supported by the Swiss National Science Foundation and by
the NCCR program {\it Materials with Novel Electronic Properties}
(MaNEP) sponsored by the Swiss National Science Foundation.

\end{document}